\documentclass{article}

\usepackage{arxiv}

\usepackage[utf8]{inputenc} 
\usepackage[T1]{fontenc}    
\usepackage{hyperref}       
\usepackage{url}            
\usepackage{booktabs}       
\usepackage{amsfonts}       
\usepackage{nicefrac}       
\usepackage{microtype}      
\usepackage{amsmath}
\usepackage{cleveref}       
\usepackage{lipsum}         
\usepackage{graphicx}
\usepackage{natbib}
\usepackage{doi}

\usepackage[most]{tcolorbox}
\usepackage{listings}
\usepackage[table]{xcolor}
\usepackage{multirow}

\newcommand{\aimigrate}{AIM\textsc{igrate}}
\newcommand{\starsim}{S\textsc{tarsim}}
\newcommand{\parcels}{P\textsc{arcels}}
\newcommand{\langchain}{L\textsc{ang}C\textsc{hain}}
\newcommand{\typhoidsim}{T\textsc{yphoidsim}}
\newcommand{\briefgpt}{B\textsc{rief}GPT}

\title{What a diff makes: automating code migration with large language models}

\newif\ifuniqueAffiliation

\ifuniqueAffiliation 
\author{ \href{https://orcid.org/0000-0000-0000-0000}{\includegraphics[scale=0.06]{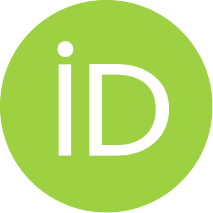}\hspace{1mm}David S.~Hippocampus}\thanks{Use footnote for providing further
		information about author (webpage, alternative
		address)---\emph{not} for acknowledging funding agencies.} \\
	Department of Computer Science\\
	Cranberry-Lemon University\\
	Pittsburgh, PA 15213 \\
	\texttt{hippo@cs.cranberry-lemon.edu} \\
	\And
	\href{https://orcid.org/0000-0000-0000-0000}{\includegraphics[scale=0.06]{orcid.pdf}\hspace{1mm}Elias D.~Striatum} \\
	Department of Electrical Engineering\\
	Mount-Sheikh University\\
	Santa Narimana, Levand \\
	\texttt{stariate@ee.mount-sheikh.edu} \\
}
\else
\usepackage{authblk}

\newbox{\orcid}\sbox{\orcid}{\includegraphics[scale=0.06]{orcid.pdf}} 
\author[1]{Katherine A. Rosenfeld}
\author[1]{Cliff C. Kerr}
\author[1]{Jessica Lundin}
\affil[1]{Institute for Disease Modeling, Gates Foundation}
\fi

\renewcommand{\headeright}{}
\renewcommand{\undertitle}{}
\renewcommand{\shorttitle}{Automating code migration with large language models}

\hypersetup{
pdftitle={A template for the arxiv style},
pdfsubject={q-bio.NC, q-bio.QM},
pdfauthor={David S.~Hippocampus, Elias D.~Striatum},
pdfkeywords={First keyword, Second keyword, More},
}

\begin{document}
\maketitle

\begin{abstract}
Modern software programs are built on stacks that are often undergoing changes that introduce updates and improvements, but may also break any project that depends upon them.  In this paper we explore the use of Large Language Models (LLMs) for code migration, specifically the problem of maintaining compatibility with a dependency as it undergoes major and minor semantic version changes. We demonstrate, using metrics such as test coverage and change comparisons, that contexts containing diffs can significantly improve performance against out of the box LLMs and, in some cases, perform better than using code. We provide a dataset to assist in further development of this problem area, as well as an open-source Python package, \aimigrate, that can be used to assist with migrating code bases. In a real-world migration of \typhoidsim\,between \starsim\,versions, \aimigrate\,correctly identified 65\% of required changes in a single run, increasing to 80\% with multiple runs, with 47\% of changes generated perfectly.
\end{abstract}


\section{Introduction}

Large Language Models (LLMs) have demonstrated impressive capability in various coding problems, including generation, refactoring, and translation \cite{roziere22, tahranijamsaz24}. In this paper we focus on a specific subtask under code translation: migrating code to maintain compatibility with major semantic version changes to a dependency. This task is common to projects involving software maintenance, application deployment, and research program management. Furthermore, it is a problem that affects teams ranging from individuals to collaborations. 

One of the challenges associated with this problem is that it requires changes that both maintain functionality and ensure compatibility with the existing (and sometimes rapidly changing) ecosystem of the project and its dependencies. This differentiates the task from many other coding problems where the problem is generally structured by the language and basic functionality. Relatedly, there have been significant results from tool calling to code generation and repair \cite{patil23, tang24, wang24, zheng24}. Simultaneously, there have been a growing number of solutions (many available only at cost) with impressive performance across tasks related to software engineering \cite{yang24}.

However, these solutions generally do not serve smaller projects where migration is an important part of maintaining communities and impact. For example, the task is not captured in current benchmarks \cite{chen21, jimenez23, shah24}. On the other hand, there are established methods to help with the migration process, including the identification and classification of refactoring patterns \cite{tsantalis18, tsantalis20}, along with the automated updates based on these identified patterns \cite{pawlak15}. However, these tools are often paired with specific libraries and languages, rather than being general purpose solutions to the migration problem itself.

In this work, we explore pairing \textit{diff} utilities and LLMs to approach this problem using a zero-shot approach that leverages the large context window of state-of-the-art models (e.g., 128k tokens for gpt-4o). A \textit{diff} utility uses algorithms that find the longest common subsequence (LCS) and creates a script for transforming one sequence into another (e.g., the LCS of ``dolphin'' and ``penguin'' is ``pin''). Diffs have been used for a variety of interesting applications from modeling the evolution of topics in software projects \cite{thomas11} to training LLMs to use APIs \cite{zan24}; however, we are not aware of diffs being used directly for code translation or generation.

\begin{figure*}[ht]
\vskip 0.2in
\begin{center}
\centerline{\includegraphics[width=\textwidth]{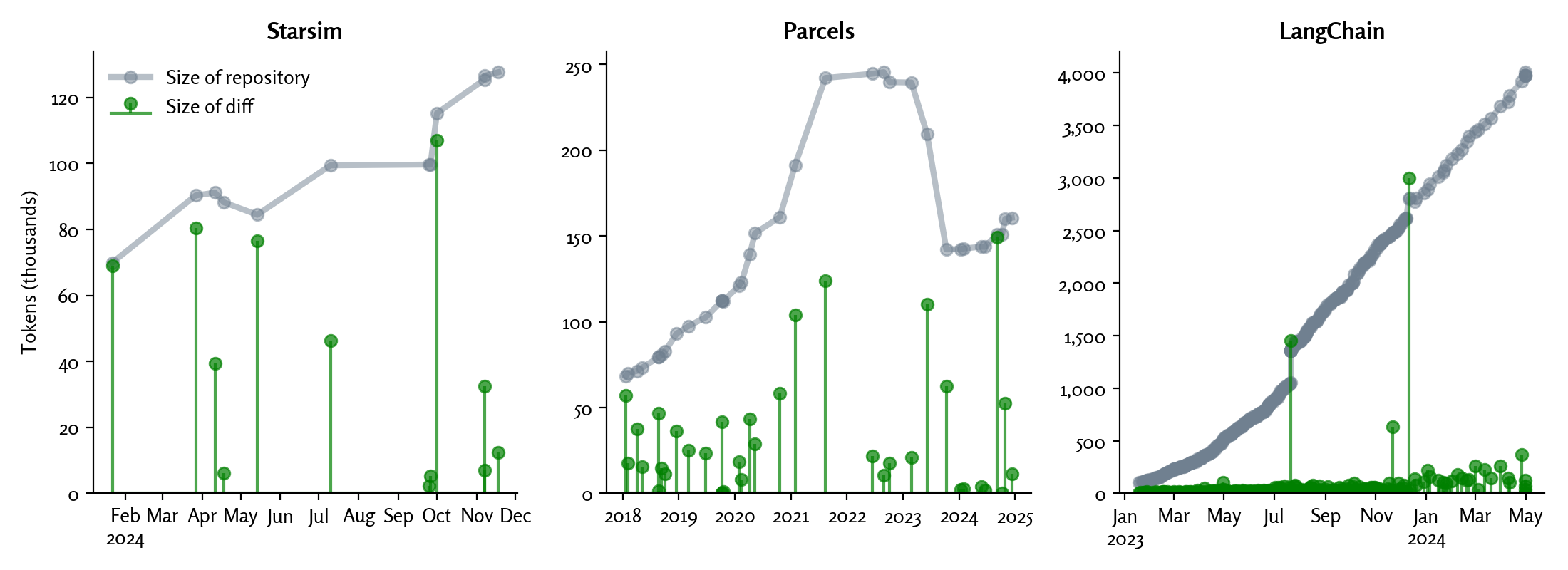}}
\caption{Sizes of the \starsim, \parcels, and \langchain\;repositories over the course of their commit history (gray). The size of the commit differences is plotted in green. We measure the size using the o200k\_base tokenizer and filter for Python files (.py) and exclude docs/ and tests/ directories.}
\label{fig:repo_sizes}
\end{center}
\vskip -0.2in
\end{figure*}

In the next section, we motivate the study by assessing the performance of LLMs for diff comprehension. Section \ref{sec:method} describes our methodology for code migration and an associated Python package, \aimigrate\footnote{https://github.com/starsimhub/aimigrate}. We investigate the capability and limitations of this approach in section \ref{sec:experiment} by exploring three case studies. Section \ref{sec:correctness} focuses further on assessing the correctness of the code. Section \ref{sec:conclusion} discusses limitations and future directions.

\section{Motivation}
\label{sec:motivation}

State-of-the-art LLMs demonstrate impressive performance out of the box on a wide range of challenging benchmarks. In practice, however, using LLMs often requires data compression through some form or combination of filtering (e.g., RAG), structuring (e.g., graphs), or tuning (e.g., fine-tuning, test-time-training). One reason we consider diffs is that they present a concise way to capture changes in code. Figure \ref{fig:repo_sizes} shows the size of three Python libraries over their commit history measured using the o200k\_base encoding. We compare the total size of the repository (filtering for \texttt{.py} files and excluding \texttt{docs} and \texttt{tests} directories) against the sequential diff commits. Notably, the size of the diff is almost always less than the size of the repository and often significantly so. Furthermore, changes to the code base are co-located by design in the diff output. So while the size of the diff will likely increase with the number of commits, it is clear that there are some cases where the diff provide some form of compression as well as concentration. 

The general concept of differencing has been utilized in multiple LLM applications including pre-training \cite{zhang22}, training for API invocation \cite{zan24}, differential testing \cite{rao24}, and code review/commit message generation \cite{fan24}. In particular, \citet{fan24} reported that pairing in context learning with diff formatted code changes resulted in better performance for comment generation than using code. However, we are not aware of quantitative, general assessments of LLM comprehension of diff outputs. In order to motivate our usage of diffs for the task at hand (code migration) we explore this issue to see a) if LLMs can comprehend diff outputs for coding tasks and b) how the quality of the comprehension compares to using pure code. 

The HumanEval dataset, released by OpenAI in 2021, was designed to test code generators using docstrings. The benchmark, consists of 165 function headers and docstrings, solutions, and tests \cite{chen21}. Using 100 of the provided solutions, we use Claude 3.5 Sonnet to generate two solutions: one solution is prompted for correctness and the other for a mistake. The prompt strategy is meant to produce meaningful changes in the style and implementation of the code. Using the harness 
provided by OpenAI, we filtered the results to ensure that code would produce the correct pass/fail status leaving 174/200 verified test questions. 

\begin{figure*}[ht]
\begin{center}
\centerline{\includegraphics[width=\textwidth]{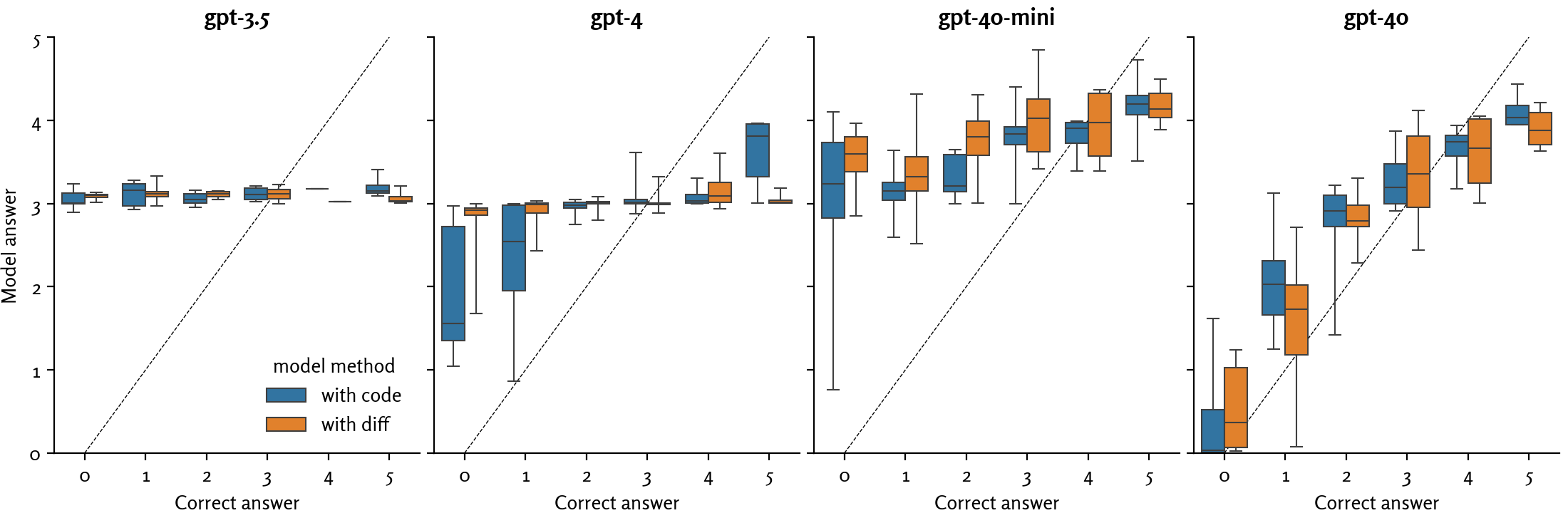}}
\caption{Results from the diff comprehension test. The figure plots the LLM output using the code (blue) versus diffs (orange) as a function of the correct answer (the number of functions with an error). Each sub-panel corresponds to a different LLM ($n=50$ test questions per LLM and method).}
\label{fig:comprehensions}
\end{center}
\vskip -0.2in
\end{figure*}

For our ``diff comprehension test'', each test question is composed a subset of three files:
\begin{itemize}
    \item FileA.py: A concatenation of 5 randomly chosen functions from the HumanEval dataset solutions (function header + function body with no docstring)
    \item FileB.py: A concatenation of the same functions using the new solutions where we uniformly choose between 0 and 5 of the functions to be incorrect
    \item FileC.py: the diff between FileA.py and FileB.py.
\end{itemize}

We assess the LLM's output count of the number of functions that have errors over two trials where the first trial includes FileA.py and  FileB.py and the second trial includes FileA.py and FileC.py. Diff comprehension can be assessed by comparing performance against random answer and the answer using pure code. Figure \ref{fig:comprehensions} plots the results for several of the OpenAI models. To account for uncertainty in the LLM output, we compute a weighted average of the token probabilities, $p(t)$: $\sum_{t=0}^{5} t p(t)$.

There are few interesting things to note from this exercise. First, figure \ref{fig:comprehensions} shows that smaller models like gpt-3.5-turbo and gpt-4o-mini perform poorly on this test. However, gpt-4o-preview and gpt-4o perform significantly better. Gpt-4-preview exhibits significantly poorer performance using the diff, as indicated by a Wilcoxon signed-rank test on the $L_1$-norm (p=0.014). Gpt-4o achieves the closest parity between the methods. This suggests that LLMs {\it can, but do not necessarily} perform well on coding tasks using diffs. Additional exploration of this issue is provided in section \ref{sec:experiment}. Scores evaluated for additional models is presented in appendix \ref{appendix:diff_scores}



\section{Method}
\label{sec:method}

\begin{figure}[ht]
\centerline{\includegraphics[width=\columnwidth]{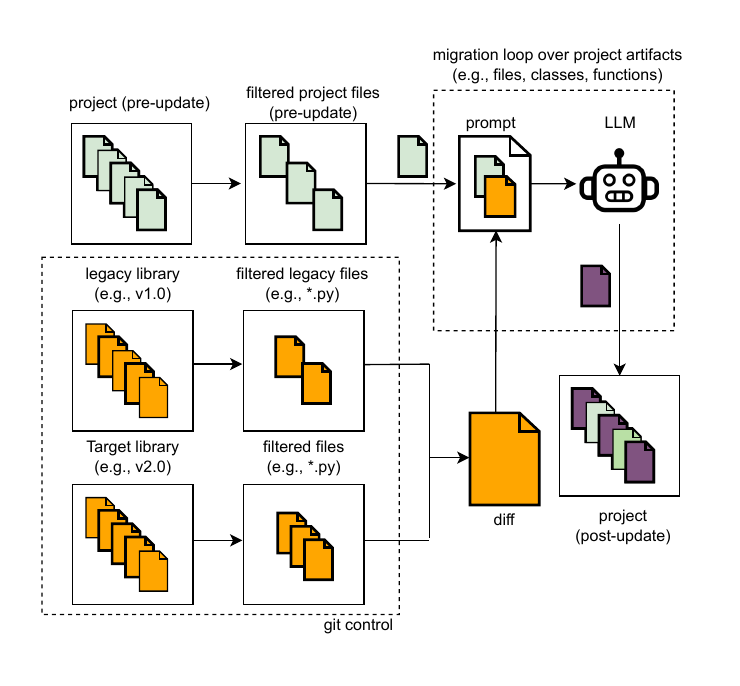}}
\caption{Diagram showing how AIM\textsc{igrate} uses diffs to automate code migration. The diagram does not capture how the pre-update project files are filtered for migration nor does it capture the quality and safety checks applied to the resulting code. The migration loop can be run multiple times to increase overall quality of the results (section \ref{sec:correctness}) Figure made with app.diagrams.net/.}
\label{fig:aimigrate}
\vskip -0.2in
\end{figure}

The migration workflow (see figure \ref{fig:aimigrate}) requires specification of 1) the legacy library version, 2) target library version, and 3) the pre-update project files. In practice, since the repositories and resulting diff files can exceed the context window size of current models, the workflow also requires filtering for for which library artifact (e.g., files) to use in the migration process. While this step could be automated using e.g. \texttt{ast}, we use file patterns to either include or exclude files. This specification is used to construct a single diff containing all the changes to the filtered files between the legacy and target versions of the library\footnote{We use the git CLI default diff implementation \cite{myers86}}.  The diff is combined with each pre-update project file, separately, and the LLM output is the post-update project file. The collection of these post-update files can then be assessed for quality and accuracy. Further details about prompt construction are provided in Appendix \ref{appendix:prompts}.

Key aspects of this approach include:
\begin{itemize}
    \item There is no dependency on the library or project code, thus avoiding any potential conflicts.
    \item While the examples presented in this paper are exclusively for Python, the approach is not limited to any specific programming language.
    \item It does not require generating specialized scripts for the dependency (cf. migration approaches based on, for example, G\textsc{rit}QL) and the same workflow can be applied to multiple libraries for a fixed project.
    \item The workflow sketch (figure \ref{fig:aimigrate}) does not include checks of the generated code; any code produced by the LLM should be inspected for both safety and quality.
\end{itemize}

We provide an implementation of this diff-based workflow (along with other approaches discussed in section \ref{sec:experiment}) as a publicly available Python library, \aimigrate. The tool supports OpenAI, Anthropic, Gemini, and local models (using ollama). See Appendix \ref{appendix:aimigrate_example} for an example.

\section{Experimental setup}
\label{sec:experiment}

Our experimental design consists of a set of experiments, $\mathcal{E}$, where each experiment $E(l, m, c) \in \mathcal{E}$ is defined by a LLM model $l$, a migration method $m$, and a case study $c$. In this section we describe the set of case studies $\mathcal{C}$ and then the metrics used to evaluate $\mathcal{E}$.

\begin{figure*}[ht]
\begin{center}
\centerline{\includegraphics[width=\textwidth]{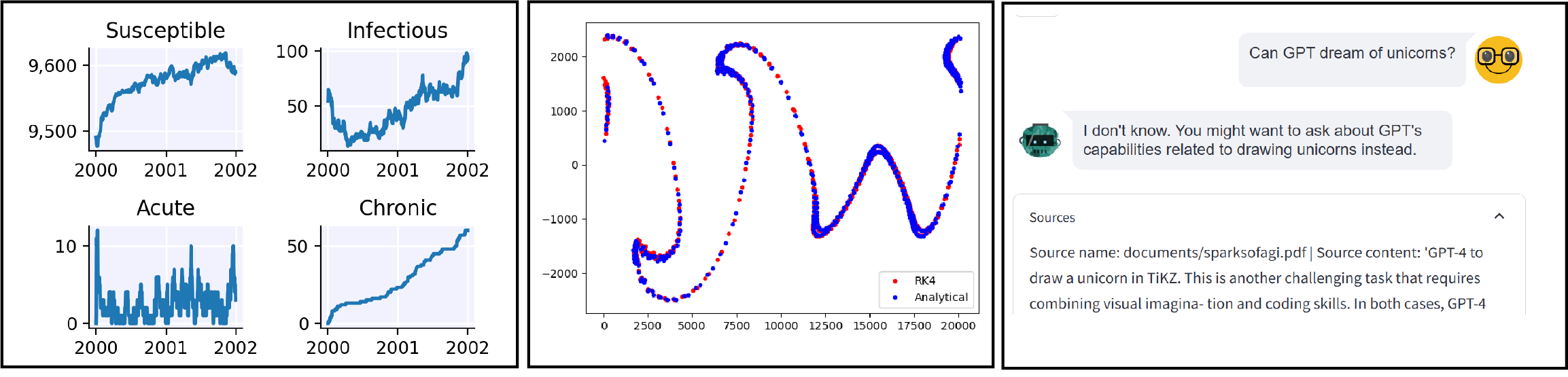}}
\caption{Figure showing examples of the case studies: time series from a \typhoidsim\ simulation (left), a comparison of advection kernels in a \parcels\ tutorial (center), and Q\&A in the \briefgpt\ app (right).}
\label{fig:examples}
\end{center}
\vskip -0.2in
\end{figure*}

\subsection{Data}

We present three case studies that highlight strengths and challenges of the methodology. Each case study contains files compatible with the legacy and target library. The post-update files are a reference created by the project maintainers. We then apply the diff-based workflow using \aimigrate, along with two other migration strategies, to the initial migration problem and compare our results to the reference post-update files by using a testing framework approach as well as via direct line-by-line comparisons. 

These case studies were selected for their diversity (ranging from software/application development to disease modeling). The include recent updates featuring changes to syntax, structure, and features. The code supplied with this analysis provides instructions for adding additional case studies. Furthermore, we provide a dataset of working, minimal changes that capture the migration task. This data is utilized to assess the diff-based approach.

Each of the case studies is based on an existing, open-source\footnote{\typhoidsim\ will available publicly upon acceptance} project on GitHub. However, the structure of each is different (e.g, library, scripts, and notebooks) so we added a formatting step to enable code assessment via a testing framework (see section \ref{sec:evaluation}).

\subsection{Evaluation}
\label{sec:evaluation}

\aimigrate\ should be paired with thorough review and assessment; in practice it provides a useful starting point to identify potential changes necessary for the migration. However, we assess the out of the box performance of the workflow on the three case studies described in detail below. We report two methods of assessment: performance on a suite of tests and line-by-line classification performance. Each case study comes with a suite of tests that assess functionality of the post-update project. We use \texttt{pytest} to automate running the tests and tracking the number that pass. The second assessment is a line-by-line classification where we first identify all significant changes to the pre-update project from our reference version (generated by a human / associated with the project). We estimate recall and precision by looking for a matching line change between the \aimigrate\ version and the reference version. These metrics assess correctness line-by-line and complement the test based approach. We conduct an additional analysis looking at code correctness in section \ref{sec:correctness}.

For each case study, we run three migration strategies:
\begin{itemize}
\item \textbf{Black box}: We include only the name, common alias (e.g., np for numpy) and versions of the legacy and target library in the prompt.
\item \textbf{With code}: We include the filtered code files from the target library in the prompt.
\item \textbf{With diff}: We include the diff for the filtered code files in the prompt.
\end{itemize}

We tried \aimigrate\ with the datasets using a suite of open and private LLMs including from OpenAI (gpt-4o, gpt-4o-mini, o1, o1-mini), Google (gemini-flash-1.5, gemini-2.0-flash-exp), llama (llama3-8b), and Anthropic (claude-sonnet-3.5). Of these, we observed that only a subset of the OpenAI models (gpt-4o, gpt-4o-mini, and o1) were appropriate for this particular analysis (see appendix \ref{appendix:model_names} for specific model release information). The Gemini models had a tendency to devolve from producing Python code to incomprehensible diffs format; o1-mini would fail on single file requests; our requests exceeded Anthropic's rate limits; and llama3's memory requirements exceeded the memory of our devices. While additional engineering on the pipeline would likely reduce these model failures, we include these differences as useful notes and will focus on the three most robust models (gpt-4o, gpt-4o-mini, and o1) for our analysis.

Our evaluation pipeline involves running code generated by the LLM. Follow basic safety procedures such as using an isolated container and secure environment when running the scripts associated with this analysis \cite{gvisor}.

\begin{figure*}[ht]
\vskip 0.2in
\begin{center}
\centerline{\includegraphics[width=\textwidth]{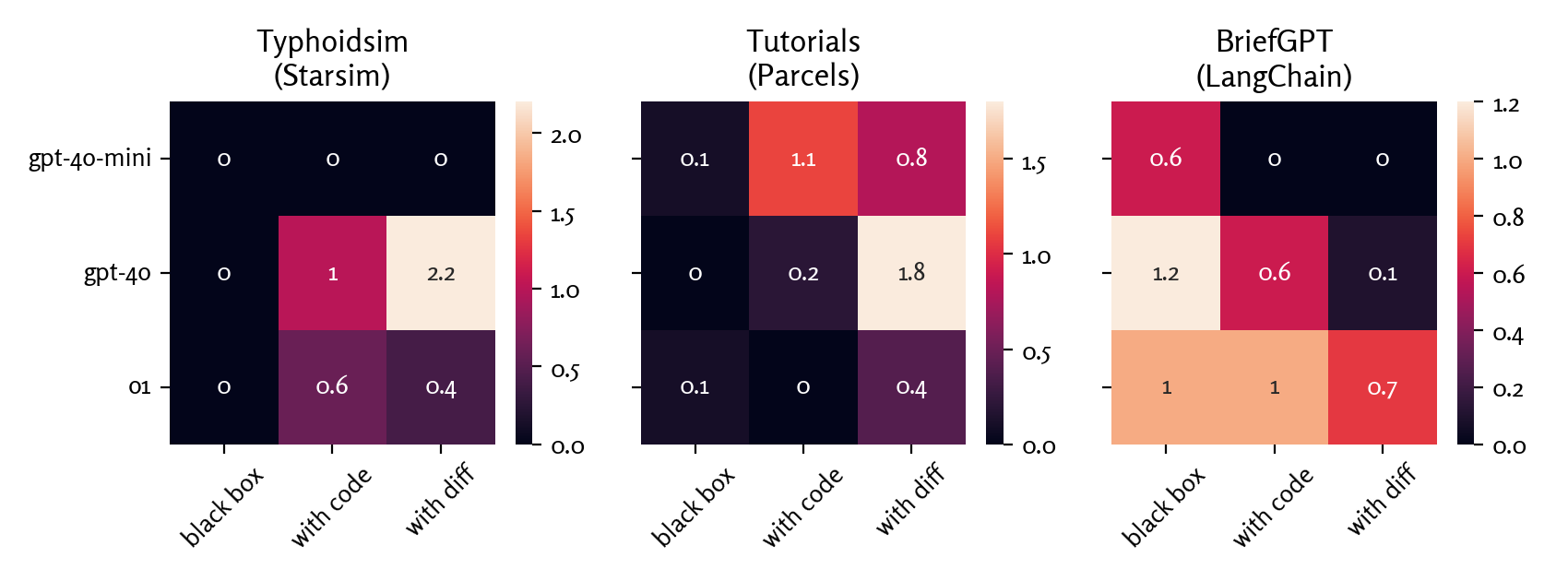}}
\caption{Heatmaps showing the mean number of tests passing by case study for each model and migration method combination.}
\label{fig:passed_heatmap}
\end{center}
\vskip -0.2in
\end{figure*}

\subsection{\starsim}
\starsim\footnote{https://github.com/starsimhub/starsim} is an agent-based modeling framework for simulating the spread of diseases (e.g., HIV, tuberculosis, cholera, Ebola) and underlying determinants of health (such as malnutrition) among agents via dynamic transmission networks \cite{starsim}. \starsim\ is a general-purpose, multi-disease framework that was built upon a pre-existing set of disease-specific models \cite{kerr23,hpvsim,fpsim}. It was first released in January 2024, with version 1.0 released in July 2024. Version 2.0 released in October 2024 and introduced several breaking changes such as module-specific timesteps, time-aware parameters, and changes to module types and integration. This example is interesting because the release of v2.0 introduced a fundamental change in how certain parameters are treated, making explicit the time dimension that had previously been implicit.

In this case study we look at the migration of the package \typhoidsim\ built with \starsim. Typhoid is a bacterial infection that is usually spread through contaminated food or water and can public health programs utilize vaccines along with water, sanitation, and hygiene (WASH)-based control. \typhoidsim\ was originally developed using v1.0.3 of \starsim, but the project was updated to use v2.2.0 at the beginning of 2025. The version (\texttt{aeb79a1}) we utilize mostly reflects an update for compatibility to \starsim\ and so provides both a relatively complex and relatively clean example (i.e., changes in the project reflect the migration itself rather than other code updates). For our assessments we do an additional round of cleaning to exclude changes to comments and additional features. An important caveat is that \aimigrate\ was used to help with the migration which was then completed by two \starsim\ maintainers.

For the migration task we supply \aimigrate\ with 6 files that contain the \starsim\ classes utilized by \typhoidsim. The nine tests are direct from \typhoidsim's testing framework (we do not include the developer tests).

\subsection{Parcels}
\parcels\footnote{https://github.com/OceanParcels/Parcels/} (\textbf{P}robably \textbf{A} \textbf{R}eally \textbf{C}omputationally \textbf{E}fficient \textbf{L}agrangian \textbf{S}imulator) is a particle tracking simulator that uses output from Ocean Circulation models \cite{lang17, delandmeter19, kehl23}. \parcels\ can be used to track both passive particles, like water, plastic, and plankton, or active like fish. The project was first published in 2017 with v0.9. The next documented release, v2.4.2, was in 2019 \cite{delandmeter19} and v3.0 followed in 2023. One of the key features of \parcels\ is a JIT mode that generates C-code to perform the particle state update for a user-defined generic kernel function. This is a common pattern used in scientific programming where Python provides a user-facing layer built over more either a more performant language. Projects such as P\textsc{y}O\textsc{pen}CL and \textsc{py}CUDA are popular examples that enable this type of project.

This case study is a migration of tutorials for \parcels\ from v2.4.2 \cite{kehl23} to v3.1.0. Among a number of changes, v3.0 introduced a critical reordering of kernel actions that enabled kernels to now be combined (e.g., advection + sinking)\footnote{https://github.com/OceanParcels/Parcels/pull/1402}. Writing kernels that can utilize this feature results in non-Python code that is flagged as compiler error by IDEs such as VScode. This presents a new challenge to the migration problem where, essentially, two code syntaxes are combined in one file. We select 3 tutorials based on their complexity and utilization of breaking changes from v2 to v3. The tutorials, maintained in Python notebooks, cover diffusion, analytical advection, and tracking data from Argo floats. We reformat the Python notebooks into scripts that can be incorporated into the testing framework. When we compare the tutorials between v2.4.2 and v3.1.0, we observe changes in content and syntax. In order to just capture the required updates to use parcels, we use the original Python notebook tutorials to set style and content.


For the migration task we supply \aimigrate\ with an example on moving eddies and a tutorial on kernel construction (2 files). We found that this latter file was critical to see any migration to the kernels to take advantage of the kernel reordering. In order to capture this, we set our tests to error when the old kernel syntax was utilized.

\subsection{\langchain}
\langchain\footnote{https://github.com/langchain-ai/langchain} is a popular framework for developing applications powered by LLMs. v0.0.1 was published in July,= 2023 and the repository had 20k commits to bring it to v0.3.14 released at the start of 2025. This project undergoes almost constant change, with over 300 published tags alone. The library provides a wide range of tools to build applications from chat bots, to summarization tools, to agentic frameworks. As a result, it can be challenging to harmonize projects using different versions of \langchain. To help with updating projects that use \langchain, there exist tools using declarative techniques e.g., G\textsc{rit}QL. 

\langchain\ began as a single library, but switched to being a framework - a collection of libraries and integration packages. This resulted in large changes in the repository structure as well as how the tools are imported into a project. The \textit{langchain-cli} can help automate these changes. Additionally, \langchain\ introduced the LangChain Expression Language (LECL) to help chain different object (``runnables'') into sequential operations. This directly affects how LLMs are called by the library.

The case study, \briefgpt, is a proof-of-concept application built using an early version of \langchain\ (v0.0.169) \cite{briefgpt}. Built using S\textsc{treamlit}, it employs two RAG approaches for text querying and summarization. Although it is not maintained, it has over 700 stars on GitHub.

The project is a collection of functions embedded in scripts and we maintain the same structure when we construct a reference migration solution to \langchain\ (v0.3.14). We write tests to evaluate tools from the \langchain\ library as they are called in \briefgpt. To focus on assessing changes in the core project code, we allow deprecated imports but treat warnings related to LECL as errors. Additionally, we provide \aimigrate\ 12 files mapping \langchain\ classes used in the code to their respective base classes.



\begin{figure*}[t]
\vskip 0.2in
\begin{center}
\centerline{\includegraphics[width=\textwidth]{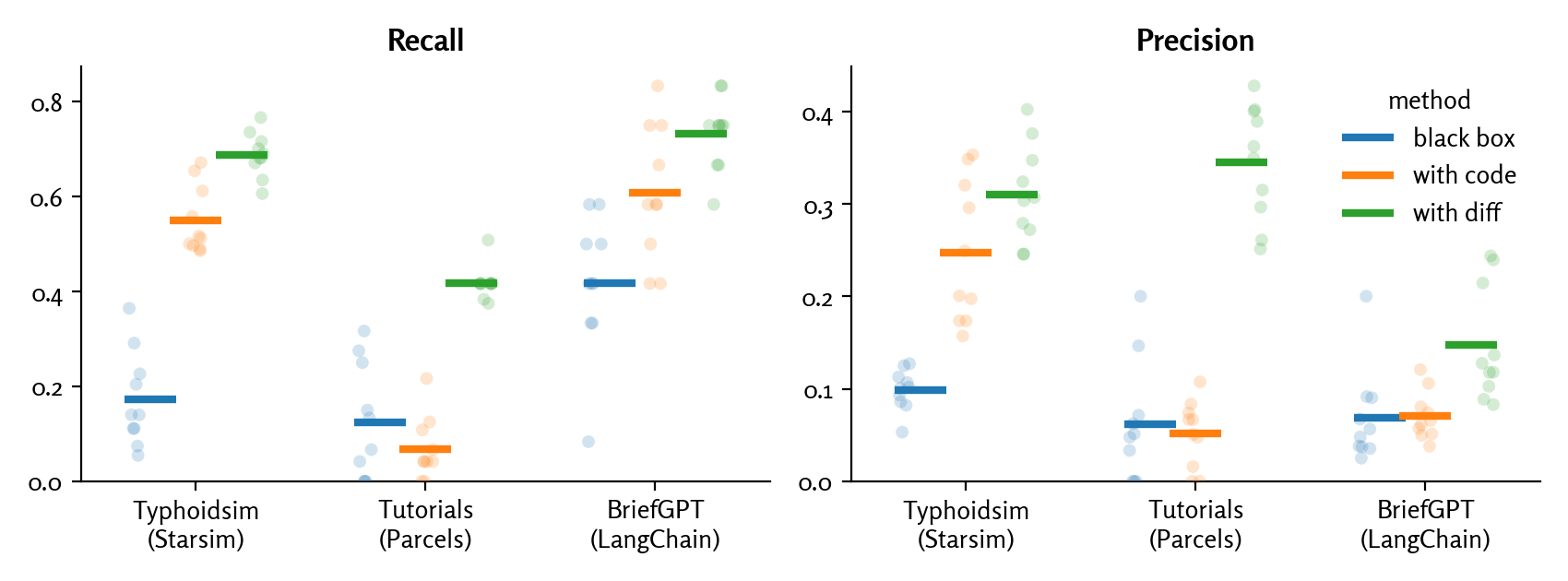}}
\caption{Recall (left) and precision (right) metrics assessed for migrations of the three case studies using gpt-4o.}
\label{fig:recall_precision}
\end{center}
\vskip -0.2in
\end{figure*}

\begin{figure*}[b]
\begin{center}
\centerline{\includegraphics[width=\textwidth]{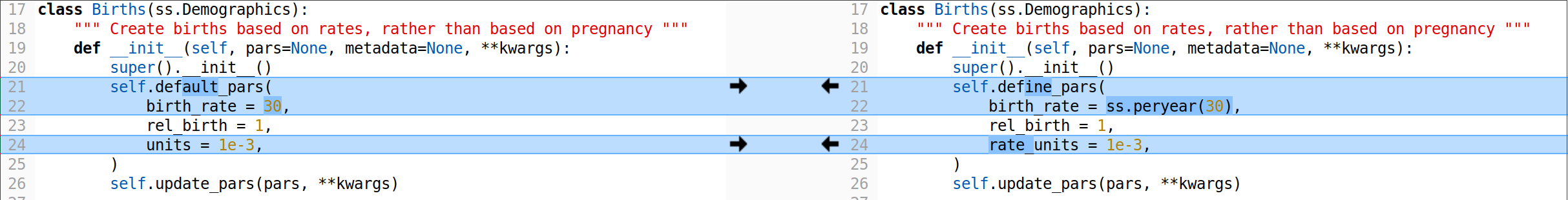}}
\caption{Example of correct changes to  \typhoidsim\ made by \aimigrate\ in migrating from \starsim\ v1.0.3 (original version, left) to v2.2.0 (reference version, right). These include renaming a method, renaming a keyword argument, and adding units to a parameter.}
\label{fig:typhoidsim_diff}
\end{center}
\vskip -0.2in
\end{figure*}

\section{Results}

We report performance metrics generated by testing frameworks and line-by-line comparisons. See section \ref{sec:evaluation} for a description of the metrics. In addition, we discuss costs associated with running the OpenAI models as well as the execution time. Each experiment, $E(m, \mu, c)$, was run 10 times and the output was corrected for small errors unrelated to code (e.g., preamble not caught by \aimigrate). 

\subsection{Unit tests}
\label{sec:unit_tests}
Figure \ref{fig:passed_heatmap} shows the mean number of tests passed, $\bar{p}(l,m|c)$. Although the overall performance is poor ($\textrm{max}\left(\bar{p}\right) = 2.2$), these represent out-of-the-box results and, in practice, the generated files must undergo inspection for quality and safety (e.g., a single syntax error anywhere in the project will cause every test to fail). However, $\bar{p}$ can also be used as an indicator of relative performance for migration method $m$ and LLM $l$. We see that for the smaller, more esoteric case studies ($c\in \left\{ \text{\typhoidsim\ and \parcels\ tutorials} \right\}$), the migration method utilizing either the code or diff tend to perform better than the black box approach, regardless of the LLM. For both of these case studies and when code was included in the context, o1 also performed worse than gpt-4o. Inspecting the errors found by the testing framework, we see that o1 commonly introduced errors during import and caused the testing framework to error during collection. 

Surprisingly, the best performance for the \briefgpt\ example came from using the LLMs as a black box (i.e., with no context). \langchain\ is popular and was established before the training cutoff of the gpt-4 models in October 2023. When prompted with ``what is langchain?'', gpt-4o-mini outputs a detailed explanation of the package (see appendix \ref{appendix:knowledge}). While \parcels\ also predates the training cutoff and can be identified by gpt-4o-mini, the release of v3.0 in October likely excludes those changes from the training data. The gpt-4o-mini does not produce an accurate description of the \starsim\ library. This case study suggests that for simple cases, such as minor changes involving popular libraries, LLMs can generate valid output without additional examples. We do not think that this is surprising - this capability is, in part, enabling the rapid development and adoption of LLM-based coding assistants and tools.

\begin{figure}[ht]
\centerline{\includegraphics[width=0.5\columnwidth]{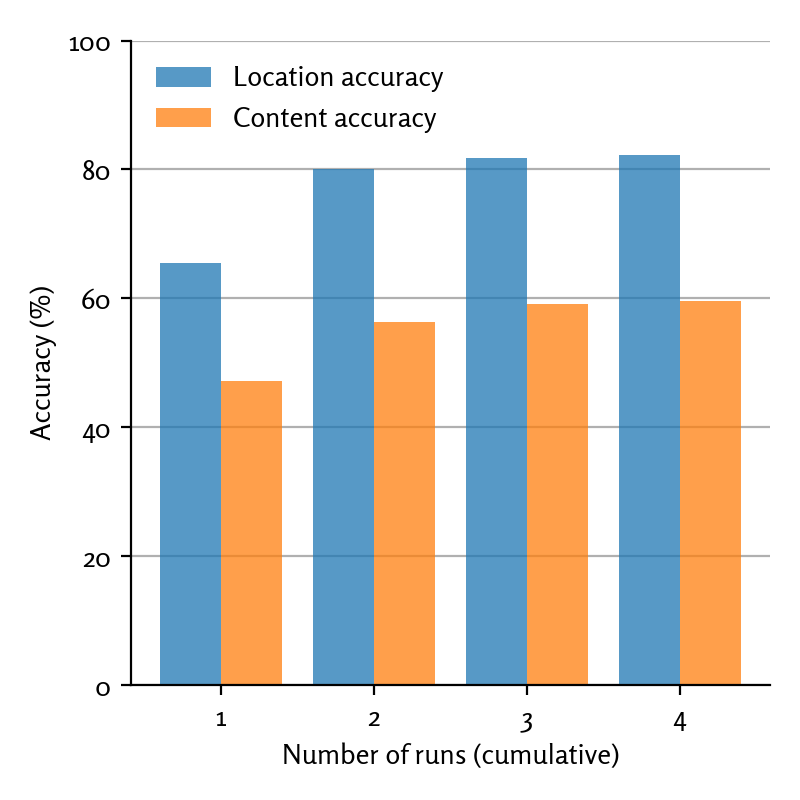}}
\caption{Accuracy of \aimigrate\ in correctly identifying the locations (line numbers) of required of changes (blue), and in correctly identifying the code changes themselves (orange). Running multiple times results in additional correct locations and code changes, although with saturation after 2-3 runs.}
\label{fig:typhoid_perf}
\vskip -0.2in
\end{figure}

\subsection{Edit matching}

Figure \ref{fig:recall_precision} plot the recall and precision for the subset of $\mathcal{E}$ using gpt-4o (see section \ref{sec:evaluation} for description of the metrics). Notably, the migrations that used the diff have consistently higher recall and precision. The difference is largest for the \parcels\ tutorials while for \typhoidsim\, utilizing the code, whether via a diff or not, rated better than an LLM out of the box. The \briefgpt\ example shows the mildest differences between the three migration methods in line with results in section \ref{sec:unit_tests}.


\subsection{Code correctness}
\label{sec:correctness}

We used \aimigrate\ in a real-world example to migrate \typhoidsim\ from \starsim\ v1.0.3 to v2.2.0. The main change was the introduction of time units, which necessitated many often subtle changes to the code (see figure \ref{fig:typhoidsim_diff} for an example), such as replacing a unitless scalar parameter value with a unit-aware object instance. Many class methods were also renamed, and some methods also had changed signatures (i.e., the new methods required different input arguments). In short, these changes were far more extensive and more complex than what could be easily accomplished through either find-and-replace or manual inspection.

We used two metrics to assess the performance of \aimigrate\ in migrating \typhoidsim\ from \starsim\ v1.0.3 to v2.2.0: accuracy of finding the \textit{location} (i.e. line number) of each required change, and accuracy of producing the \textit{correct} change. In both cases, we are treating the (AI-assisted) human-migrated code as the reference. (Note: even though there are three \textit{semantic} changes shown in the example shown in figure \ref{fig:typhoidsim_diff}, for the purposes of quantification, this would be counted as two changes with locations 21 and 24, since consecutive lines are treated as a single change by the diff algorithm.)

Although there were many cases where \aimigrate\ produced an ``almost'' correct solution (such as subtle changes to whitespace or comments), here we are counting a code change as ``correct'' if and only if it is identical to the human-migrated change.

To perform the reference migration (AI-assisted human), \aimigrate\ was run on \typhoidsim\ four times, generating four copies of the \typhoidsim\ source files. These were then manually merged with each other by a \typhoidsim\ project maintainer, keeping only valid changes in each merge. The merged version was then manually edited until the code was fully compatible with \starsim\ v2.2.0.

Performance of \aimigrate\ is shown in figure \ref{fig:typhoid_perf}. Of the 220 required changes between the original \typhoidsim\ code and the reference migration, a single run of \aimigrate\ correctly identified 144 (65\%) of these changes. The second run identified slightly fewer correct changes (140), but, since the changes identified were only partially overlapping, when combined with the first run, significantly increased the number of changes identified to 176 (80\%). Further runs resulted in few additional changes identified.

Of the 220 required changes, the first \aimigrate\ run correctly generated 104 (47\%) of them, which increased to 124 (56\%), 130 (59\%), and 131 (59\%) correct changes when combined with the three subsequent runs. Thus, even for a complex migration involving a large number of changes to code logic and syntax, \aimigrate\ was able to make nearly half of these changes perfectly from a single run.

\section{Conclusion}
\label{sec:conclusion}

Diffs provide a way to compress and localize code changes and have the potential to increase the performance of LLM-based tools. Indeed, this has already been reported for some tasks \cite{fan24}. With our diff comprehension test we provide additional evidence that some LLMs can use diffs to solve a problem with similar performance to how it would with the entire code (see section \ref{sec:motivation}). Building upon this result, we explore the potential application of diffs to help automate code migration: taking a project and updating it to use a new version of a dependency.

We present a Python library, \aimigrate, that can help automate code migration using this diff-based approach, referencing library files, or using an LLM out of the box. Using \aimigrate\ we explore three case studies drawn from three different domains: disease transmission modeling, particle tracking for ocean circulation models, and building LLM based applications. Furthermore, the structures and challenges associated with the case studies are also quite different. This diversity contained in these examples allow us to highlight where this diff-based approach may exceed, or at least have parity with, the code based approach. In addition we present a real-world example where the AI assists the human and demonstrate that the diff based approach can make nearly half of the required changes perfectly from a single run (see section \ref{sec:correctness}). Future work includes assessing different algorithms for constructing the diff (e.g., Myers vs patience) as there have been measurable effects in previous diff applications \cite{nugroho19} and improving the filtering and artifact creation for the diff.

There are several important limitations to this method. First, the selection of files to include from the library is user-specified rather than automated or rule-based. In practice the user should have knowledge of both the project and the library and so already has some prior. Second, the diffs can easily become large, exceeding the context window length of most LLMs and potentially reducing the quality of the output \cite{li24, levy24}. See appendix \ref{appendix:costs} and \ref{appendix:time} for cost and timing reports. Such a large context window ($\geq 128k$ tokens) may be prohibitive to deploy locally. 

We hope that \aimigrate\ can make a difference for migration or exploration using diff-based approaches.

\section{Disclosures}
This work is based on research conducted by the Institute for Disease Modeling, a research group within, 
and solely funded by, the Gates Foundation. The funders had no role in the analysis, or interpretation of the data; the preparation, review, or
approval of the report; or the decision to submit the manuscript for publication. All authors are 
employees of the Gates Foundation, however, this study does not necessarily represent the views of 
the Gates Foundation.

\bibliography{llm_migrate}
\bibliographystyle{mlsys2025}
\newpage
\appendix
\onecolumn

\section{Code and data availability}
\label{appendix:artifacts}
    All code and data is available under MIT licenses. Code for the paper: \url{https://github.com/krosenfeld/aimigrate_report}. 

\section{Diff comprehension scores}
\label{appendix:diff_scores}
Below are the scores for the models tested with the 5 question diff comprehension test:

\begin{minipage}{\textwidth}
\centering
\begin{tabular}{|l|r|r|r|r|} \hline
 & tested & algorithm & MAE & accuracy \\ \hline
0 & o1-2024-12-17 & default & 0.140 & 0.920 \\ \hline
1 & o1-mini-2024-09-12 & baseline & 0.140 & 0.900 \\ \hline
2 & o1-2024-12-17 & baseline & 0.220 & 0.860 \\ \hline
3 & o1-mini-2024-09-12 & default & 0.220 & 0.820 \\ \hline
4 & deepseek-r1:70b & baseline & 0.571 & 0.633 \\ \hline
5 & deepseek-r1:70b & default & 0.604 & 0.562 \\ \hline
6 & gpt-4o-2024-08-06 & baseline & 0.540 & 0.500 \\ \hline
7 & gpt-4o-2024-08-06 & default & 0.700 & 0.320 \\ \hline
8 & gpt-4-0125-preview & baseline & 1.000 & 0.320 \\ \hline
9 & llama3.3:70b & baseline & 1.100 & 0.300 \\ \hline
10 & gpt-4-0125-preview & default & 1.260 & 0.300 \\ \hline
11 & deepseek-r1:8b & default & 1.319 & 0.298 \\ \hline
12 & deepseek-r1:8b & baseline & 1.160 & 0.280 \\ \hline
13 & gpt-3.5-turbo-0125 & default & 1.360 & 0.280 \\ \hline
14 & llama3.3:70b & default & 1.280 & 0.280 \\ \hline
15 & deepseek-coder-v2 & baseline & 1.447 & 0.263 \\ \hline
16 & llama3:70b & baseline & 1.260 & 0.260 \\ \hline
17 & llama3:70b & default & 1.380 & 0.240 \\ \hline
18 & gpt-3.5-turbo-0125 & baseline & 1.480 & 0.220 \\ \hline
19 & llama3:8b & default & 1.720 & 0.220 \\ \hline
20 & llama3:8b & baseline & 1.620 & 0.220 \\ \hline
21 & claude-3-5-sonnet-20241022 & baseline & 1.560 & 0.180 \\ \hline
22 & claude-3-5-sonnet-20241022 & default & 1.560 & 0.180 \\ \hline
23 & gpt-4o-mini-2024-07-18 & baseline & 1.520 & 0.140 \\ \hline
24 & gpt-4o-mini-2024-07-18 & default & 1.680 & 0.140 \\ \hline
\end{tabular}
\end{minipage}

\section{Resources}

\subsection{LLM versions}
\label{appendix:model_names}
Here are the specifications for the models utilized in the paper:

\begin{minipage}{\textwidth}
\centering
\begin{tabular}{|l|l|} \hline
\textbf{short name} & \textbf{version} \\ \hline
    gpt-4o & gpt-4o-2024-08-06 \\ \hline
    gpt-4o-mini & gpt-4o-mini-2024-07-18 \\ \hline
    gpt-3.5 & gpt-3.5-turbo-0125 \\ \hline
    gpt-4 & gpt-4-0125-preview \\ \hline
    o1 & o1-2024-12-17 \\ \hline
    o1-mini & o1-mini-2024-09-12 \\ \hline
    claude-3.5-sonnet & claude-3-5-sonnet-20241022 \\ \hline
\end{tabular}
\end{minipage}

\subsection{Hardware}
{\bf Local models:}
For this analysis we had access to a VGPU (NVIDIA A40-48Q) with 43.9 GiB available memory (value reported with ollama running). Given the large context window requirements, ($\simeq128k$\,tokens) this was was not sufficient to run local models such as llama3-70b. We tried running smaller models (e.g., llama3-8b), but the results were not of sufficient quality to include in section \ref{sec:experiment}. 

{\bf Anthropic:}
Rate limits (80k tokens/minute) for a paid subscription to the Anthropic API was not sufficient to handle some of the larger analysis challenges.

\subsection{Cost estimates}
\label{appendix:costs}
We report cost estimates associated with the migration case studies. Token usage is tracked using \langchain's OpenAICallbackHandler (v0.3.14) and costs are from \url{https://openai.com/api/pricing/} (January 28, 2025).

\begin{minipage}{\textwidth}
\centering
\begin{tabular}{|r|r|r|r|r|r|} \hline
\textbf{Model/LLM} & \textbf{Case Study} & \textbf{Method} & \textbf{Prompt} & \textbf{Completion} & \textbf{Cost (USD)} \\ \hline
gpt-4o & BriefGPT & aim-lc & 1380716 & 34087 & 3.79 \\ \hline
gpt-4o & BriefGPT & aim-nc & 33650 & 32741 & 0.41 \\ \hline
gpt-4o & BriefGPT & aim-rc & 1229990 & 33582 & 3.41 \\ \hline
gpt-4o & Parcels & aim-lc & 2026580 & 61948 & 5.69 \\ \hline
gpt-4o & Parcels & aim-nc & 57500 & 58901 & 0.73 \\ \hline
gpt-4o & Parcels & aim-rc & 1133820 & 61551 & 3.45 \\ \hline
gpt-4o & TyphoidSim & aim-lc & 4553950 & 296218 & 14.35 \\ \hline
gpt-4o & TyphoidSim & aim-nc & 340510 & 286682 & 3.72 \\ \hline
gpt-4o & TyphoidSim & aim-rc & 3893710 & 284311 & 12.58 \\ \hline
gpt-4o-mini & BriefGPT & aim-lc & 1380716 & 33787 & 0.23 \\ \hline
gpt-4o-mini & BriefGPT & aim-nc & 33650 & 34896 & 0.03 \\ \hline
gpt-4o-mini & BriefGPT & aim-rc & 1229990 & 34614 & 0.21 \\ \hline
gpt-4o-mini & Parcels & aim-lc & 2026580 & 64437 & 0.34 \\ \hline
gpt-4o-mini & Parcels & aim-nc & 57500 & 59402 & 0.04 \\ \hline
gpt-4o-mini & Parcels & aim-rc & 1133820 & 63170 & 0.21 \\ \hline
gpt-4o-mini & TyphoidSim & aim-lc & 4553950 & 259307 & 0.84 \\ \hline
gpt-4o-mini & TyphoidSim & aim-nc & 340510 & 251355 & 0.20 \\ \hline
gpt-4o-mini & TyphoidSim & aim-rc & 3893710 & 243482 & 0.73 \\ \hline
o1 & BriefGPT & aim-lc & 1379840 & 112335 & 27.44 \\ \hline
o1 & BriefGPT & aim-nc & 33620 & 103414 & 6.71 \\ \hline
o1 & BriefGPT & aim-rc & 1230590 & 102882 & 24.63 \\ \hline
o1 & Parcels & aim-lc & 2026540 & 150325 & 39.42 \\ \hline
o1 & Parcels & aim-nc & 62860 & 239545 & 15.32 \\ \hline
o1 & Parcels & aim-rc & 1133780 & 155560 & 26.34 \\ \hline
o1 & TyphoidSim & aim-lc & 4553890 & 431103 & 94.17 \\ \hline
o1 & TyphoidSim & aim-nc & 340450 & 510909 & 35.76 \\ \hline
o1 & TyphoidSim & aim-rc & 3893650 & 443013 & 84.99 \\ \hline
\end{tabular}
\end{minipage}

\subsection{Execution time}
\label{appendix:time}
We use a timer implemented in \aimigrate\ to estimate the execution time of each migration run. The time does not include assembling the artifacts (e.g., diff), but for our case studies this was a fraction of the overall runtime. The majority of the time is spent by the code generation.

\begin{figure}[ht]
\vskip 0.2in
\begin{center}
\centerline{\includegraphics[width=\columnwidth]{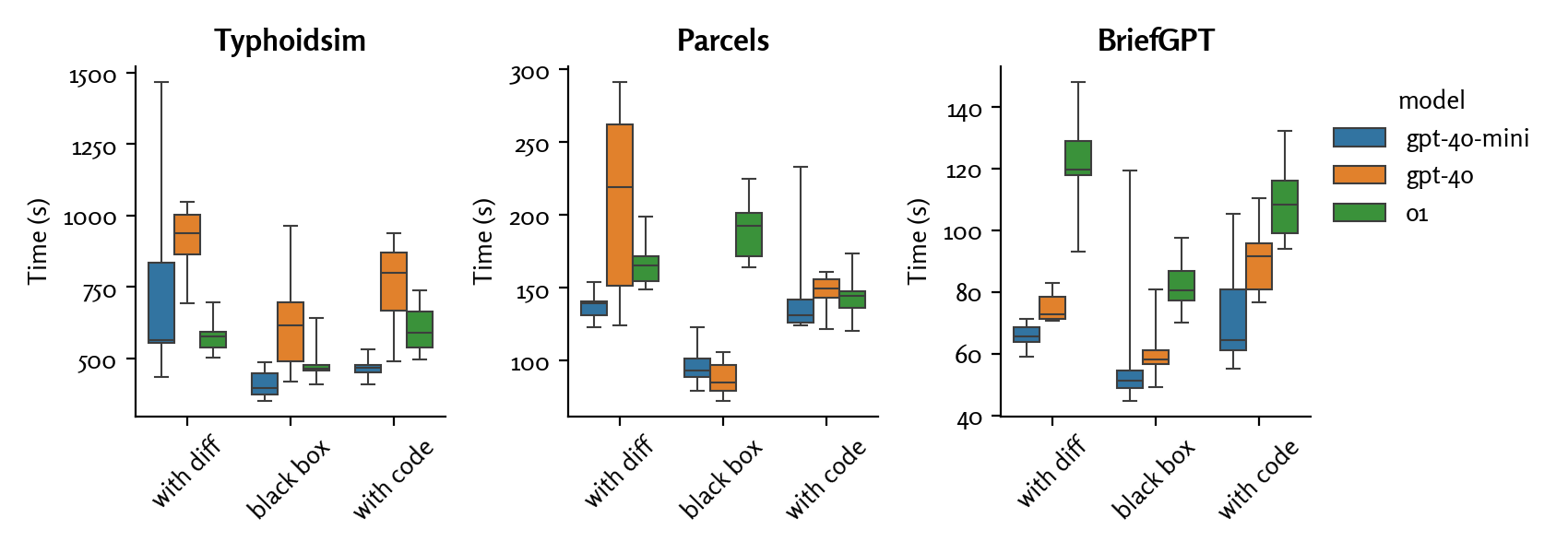}}
\caption{Box-plots showing the distribution of approximate \aimigrate\ run times for each of the three case studies by method and model ($n=10$).}
\label{fig:typhoidsim_coverage_heatmap}
\end{center}
\vskip -0.2in
\end{figure}

\section{Prompts}
\label{appendix:prompts}
\definecolor{back-blue}{cmyk}{0.34, 0.11, 0.0, 0.01, 1.00}
\definecolor{lightgray}{rgb}{.7,.7,.7}
\newsavebox\lstbox

\subsection{Diff comprehension test}
See section \ref{sec:motivation} for detailed description of the test.\\
Trial 1 (asses with code):
\begin{tcolorbox}[colframe=black, colback=back-blue, coltitle=black, title=]

\textbf{System Prompt:} You are an expert software engineer. Your task is to compare changes to functions and count how many functions now have errors.\\
\smallskip

\textbf{User Prompt:}

Here is a python file with 5 functions.

{\tt ```python\\
\{original file\}\\
```
}

Here is a python file with the same 5 functions written by someone else, but now there may be errors.

{\tt ```python\\
\{modified file\}\\
```}

Your task is to count how many of the functions now have at least one error in them. Answer with 
ONLY the number of functions that have errors [0-5].

\end{tcolorbox}

Trial 2 (asses with diff):
\begin{tcolorbox}[colframe=black, colback=back-blue, coltitle=black, title=]

\textbf{System Prompt:} You are an expert software engineer. Your task is to compare changes to functions and count how many functions now have errors.\\
\smallskip

\textbf{User Prompt:}

Here is a python file with 5 functions.

{\tt ```python\\
\{original file\}\\
```
}

Here is a diff file comparing the original file to a another file with the same 5 functions written by someone else, but now there may be errors:

{\tt ```python\\
\{diff file\}\\
```}

Your task is to count how many of the functions now have at least one error in them. Answer with 
ONLY the number of functions that have errors [0-5].

\end{tcolorbox}

\subsection{\aimigrate\ prompts}
Prompts for \aimigrate\ are not hard-coded and be passed into the \textit{Migrator} class. Below are example prompts utilized for this project:

\textbf{Black-box approach}:
\begin{tcolorbox}[colframe=black, colback=back-blue, coltitle=black, title=]

Refactor the code below to work with \texttt{\{library\}} (\texttt{\{library alias\}}). Currently the 
code works with version \texttt{\{legacy version\}} but needs to be updated to work with version \texttt{\{target version\}}.
Maintain the same style, functionality, and structure as the original code.\\
\\
\{code\}\\
\\
Refactored code:

\end{tcolorbox}

\textbf{Approach with code}:
\begin{tcolorbox}[colframe=black, colback=back-blue, coltitle=black, title=]

Below is some code for the \texttt{\{library\}} (\texttt{\{library alias\}}) library:\\
\\
\texttt{\{library code\}}\\
\\
Please refactor the code below to be compatible with the \texttt{\{library\}} library.\\
Maintain the same style, functionality, and structure as the original code.\\
\\
\texttt{\{code\}}\\
\\
Refactored code:

\end{tcolorbox}

\textbf{Approach with diff}:
\begin{tcolorbox}[colframe=black, colback=back-blue, coltitle=black, title=]

Here is the diff information for an update to the  \texttt{\{library\}} (\texttt{\{library alias\}}) library:\\
\texttt{\{ diff\}}\\

Please refactor the code below to maintain compatibility with the {library} library.\\
Maintain the same style, functionality, and structure as the original code.\\
\\
\texttt{\{code\}}\\
\\
Refactored code:

\end{tcolorbox}

\section{Knowledge of case study libraries}
\label{appendix:knowledge}
Answers are from gpt-4o-mini (training cutoff October 2023)

\textbf{What is the starsim python package?}\\
The `starsim` Python package is a tool designed for simulating star formation and evolution in astrophysical contexts. While I don't have specific up-to-date details on the package, its general purpose is to provide a framework for modeling and analyzing the dynamics of stars, including their formation, evolution, and interactions within a galactic environment.

This package is likely to include functionalities such as:

- Simulating the life cycle of stars from birth in molecular clouds to their eventual end states (e.g., white dwarfs, neutron stars, black holes).
- Modeling various physical processes involved in star formation, including gravitational collapse, nuclear fusion, and stellar winds.
- Analyzing the effects of different initial conditions and parameters on star formation rates and stellar population distributions.

To find the most accurate and up-to-date information, including installation instructions, features, and usage examples, you should check the official documentation or the repository where `starsim` is hosted, like GitHub or the Python Package Index (PyPI).

\textbf{What is the parcels python package?}\\
The `parcels' Python package is a tool designed for simulating the transport and dispersion of particles in oceanic and other fluid environments. It is particularly useful in the fields of oceanography, environmental science, and computational fluid dynamics. The package provides a framework for defining particle properties, movement equations, and the interactions of particles with their surrounding environment based on various flow models.

Key features of the Parcels package include:

1. **Particle Tracking**: Simulates the movement of particles based on flow fields, which can be derived from various sources, including observational data or numerical models.

2. **Flexible Framework**: Users can customize the behavior of particles, including their initial conditions, advection schemes, and interactions with environmental factors.

3. **Integration Methods**: Supports multiple numerical integration methods for the advection of particles, allowing users to choose the most appropriate method for their specific application.

4. **Compatibility**: Works with various data formats and can be integrated with other oceanographic tools and libraries in Python, such as NumPy and xarray.

5. **Visualization**: Provides utilities for visualizing the trajectories of particles and analyzing the results of simulations.

Parcels is beneficial for researchers and practitioners looking to model and understand the dynamics of particles in various fluid systems, including ocean currents, estuarine mixing, and pollutant dispersion.

To get started with Parcels, users typically need to install it via pip and then refer to the documentation for details on how to set up and run simulations. The package is actively maintained and has a growing community of users contributing to its development and application in various scientific studies.

\textbf{What is langchain?}\\
LangChain is a framework designed for developing applications powered by language models. It provides a modular and flexible approach to building applications that leverage the capabilities of large language models (LLMs) such as OpenAI's GPT models. The framework is particularly useful for creating conversational agents, chatbots, and other applications that require natural language understanding and generation.

Key features of LangChain include: \\
1. **Modularity**: LangChain allows developers to compose various components, such as prompt templates, memory, and chains, to create complex workflows.

2. **Chains**: These are sequences of calls to language models, tools, or APIs that can be combined to perform more complex tasks.

3. **Memory**: LangChain supports features that allow applications to remember past interactions, which can enhance user experience and contextual understanding.

4. **Integration with APIs and Tools**: LangChain can easily integrate with various external services, databases, and APIs, allowing developers to build more sophisticated applications.

5. **Customization**: Developers can customize prompts, manage contexts, and use different languages and modalities to tailor the application's behavior.

LangChain is especially useful for developers looking to create applications that require dynamic interaction, context management, and the ability to handle complex queries using natural language processing. It caters to a wide range of use cases, from simple chatbots to more advanced AI-driven applications across different industries.

\newpage
\section{Using \aimigrate}
\label{appendix:aimigrate_example}
Below is an example script running the \aimigrate\ workflow for \typhoidsim:

\begin{tcolorbox}[colframe=black, colback=lightgray, coltitle=black, title=]
\begin{lstlisting}[language=Python]
import starsim as ss
import aimigrate as aim

# The files to migrate
files = [
    'typhoid.py', # Put this first, most likely to fail
    'environment.py',
    'demographics.py',
    'interventions.py',
    'monitors.py',
    'networks.py',
    'patterns.py',
    'utils.py',
]
# Files to exclude from Starsim
exclude = [
    "docs/*",
    "tests/*",
    "__init__.py",
    "setup.py",
    'starsim/networks.py',
    'starsim/distributions.py',
    'starsim/diseases/syphilis.py',
]

M = aim.Migrate(
    source_dir = '../typhoidsim', # folder with the code to migrate
    dest_dir = './migrated_new2', # folder to output migrated code into
    files = files, # the specific files to migrate
    library = ss, # can also be the path to a cloned repository
    v_from = 'v1.0.3', # can be any valid git tag or hash
    v_to = 'v2.2.0', # ditto
    exclude = exclude,
    model = 'gpt-4o', # see aim.Models for list of allowed models
    parallel = True,
    die = False,
)
M.run()
\end{lstlisting}
\end{tcolorbox}

\end{document}